\documentclass[a4wide,12pt]{article}
\usepackage[margin=1.2in]{geometry}
\usepackage[utf8]{inputenc}
\usepackage[english]{babel} 
\usepackage{hyperref}
\usepackage{amsmath,amssymb,latexsym}
\usepackage{slashed}
\usepackage{graphicx}
\usepackage{epstopdf}
\usepackage[sorting=none, backend=bibtex,bibencoding=utf8]{biblatex}
\usepackage{mathtools}
\bibliography{ref-2}

\def \sec{\begin{section}}
\def \esec{\end{section}}

\def \La {\Lambda}
\def \la {\lambda}

\def \lc {\mathcal{L}}

\def \pr {\partial}
\def \ra {\rightarrow}

\def \beq { \begin{equation}}
\def \eeq {\end{equation}}

\DeclareMathOperator*{\tr}{tr}

\newcommand\const{\operatorname{const}}

\def \l {\left(}
\def \r {\right)}

\def \uv {\text{uv}}

\def \si {\sigma}

\def \cb {\mathbb{C}}

\def \cp {\cb P}

\begin{document}

\vskip 2cm

\begin{center}
{\Large \bf
$\cp^N$ sigma model on a finite interval revisited
}

\vskip 1.0cm

A. Milekhin\footnotetext{email: milekhin@princeton.edu}
\bigskip
\bigskip

\begin{tabular}{c}
Department of Physics, Princeton University, 08540, Princeton, NJ\\
Institute for Information Transmission Problems of Russian Academy of Science, \\
B.~Karetnyi 19, Moscow 127051, Russia\\
Institute of Theoretical and Experimental Physics, B.~Cheryomushkinskaya 25,\\
Moscow 117218, Russia
\end{tabular}

\vskip 1cm

\textbf{Abstract}
\end{center}

\medskip
\noindent
In this short note we will revisit the large $N$ solution of $\cp^N$ sigma model on a finite interval of length $L$. We will find 
a family of boundary conditions for which the large $N$ saddle point can be found analytically.
For a certain choice of the boundary conditions the theory has only one phase for all values of $L$. Also, we will provide an example when there are two phases: for large $L$ there is a standard phase with an unbroken $U(1)$ gauge symmetry and for small $L$ there is Higgs phase with a broken gauge symmetry. 

\bigskip

\section{Introduction}

Two dimensional $\cp^N$ sigma model in the large $N$ limit was first solved in \cite{adda} and \cite{witten}. 
The theory exhibits a plethora of non-trivial properties: asymptotic freedom, confinement and dynamical scale
$\La$ generation via the dimensional transmutation:
\beq 
\label{dyn}
\La^2 = \La_{\uv}^2 \exp \l - \dfrac{4 \pi}{g^2} \r 
\eeq
where $g$ is the coupling constant.

Physically, 2D $\cp^N$ model naturally arises as a low-energy effective action of non-Abelian strings in QCD-like models, see
\cite{solitons} for a review. Therefore, a finite interval geometry corresponds to a string stretched between two branes or a monopole--anti-monopole pair. Such configuration was studied in \cite{boundary}.

Recently $\cp^N$ sigma model on a 
finite interval of length $L$ with Dirichlet boundary conditions(BC) was investigated in \cite{old} and \cite{kon} using 
large $N$ expansion. 
In the earlier work \cite{old} the large $N$ saddle point equations were solved only approximately and two distinct phases were found. In \cite{kon} saddle-point equations
were solved numerically and it was argued that there is only one phase. In this paper we will find a set boundary conditions for which the saddle point equations can
be solved analytically. Strictly speaking, we will study $\cp^{2N+1}$ sigma model. We will
consider two different boundary conditions: 
\begin{itemize}
\item Mixed Dirichlet-Neumann(D-N) boundary conditions which will break global $SU(2N+1)$ to
$SU(N) \times SU(N)$. We will show that the system has at least two phases: for
$L>\pi/4\La$ there is a standard "Coulomb" phase with an unbroken $U(1)$ gauge symmetry. This phase takes place for the $\cp^N$ model on usual $\mathbb{R}^2$. For $L<\pi/4\La$ there is "Higgs" phase with broken $U(1)$. Global $SU(N)\times SU(N)$ stays unbroken in both phases.
\item Dirichlet-Dirichlet and Neumann-Neumann(D-D and N-N) boundary conditions which will break $SU(2N+1)$ to $SU(N)\times SU(N+1)$. In this case, for all values of $L$ there is a standard phase with an unbroken $U(1)$ gauge symmetry. Higgs phase is prohibited in this case, because
it will break global $SU(N)\times SU(N+1)$ to $SU(N)\times SU(N)$.
\end{itemize}
In case of simple Dirichlet boundary conditions studied in \cite{old,kon}, Higgs phase does
not break any global symmetries, so we expect that the system will have two phases as was predicted in \cite{old}.
Let us note that the large $N$ $\cp^N$ model on a cylinder also possesses multiple phases \cite{finite}.

\section{Generalized saddle point equations}
Let us study $\cp^{2N+1}$ model in the large $N$ limit. The field content consists of $2N+1$ fields $n^i, \ i=0,\ldots,2N$, real vector field $A_\mu$ and real scalar $\la$. In the Euclidian space the Lagrangian reads as:
\beq
\lc = (D_\mu n^i)^* (D^\mu n^i) +\la(n^{*i} n^i-r) 
\eeq
where $D_\mu = \pr_\mu - i A_\mu, \ \mu=t,x$ and $r=2N/g^2$. 
Time coordinate $t$ takes values from $-\infty$ to $+\infty$ and $x \in [0,L]$.

Non-dynamical Lagrangian multipliers $A_\mu$ and $\la$ forces $n^i$ to lie on $\cp^{2N+1}$ space: integration over $\la$ yields $\sum_i n^{*i} n^i = r$ and $A_\mu$ is responsible for $U(1)$ invariance $n^i \sim e^{i \phi} n^i$.

We will proceed in a standard fashion: we will integrate out 2N fields $n^i, \  i=1,\ldots,2N$ fields and then find the large N saddle point values of $\la$, $A_\mu$ and the remaining $n^0$ which we will denote by $\si=n^0$.
After integrating out $2N$ $n^i,$ fields
we have:

\beq
S_{eff} =\tr \log(-D_x^2-D_t^2+\la) + \int \ d^2x \l (D \si)^2 + \la(|\si|^2-r) \r
\label{trlog}
\eeq 
So far we do not have a factor of $2N$ in front of the 
determinant because we will impose different
boundary conditions for these $2N$ fields.

We will study this model on a finite interval of length $L$ with various boundary conditions. Note that the translational symmetry
in $x$ direction is explicitly broken. However, we still have the time translations so we will consider only time translation
invariant saddle points.
By the choice of gauge we can always set $A_t=0$. 
This allows us to rewrite eq. (\ref{trlog}) as:
\beq
\label{sgood}
S_{eff} =\sum_n E_n + \int \ d^2x \l (D_x \si)^2 + \la(|\si|^2-r) \r
\eeq 
Note that we have already integrated out time frequencies, so we have
energies $E_n$ instead of of the usual $\log \det$.
The sum over $n$ is the sum over the eigenvalues $E_n^2$ of the following equation:
\begin{equation}
(-D_x^2+\la(x))\psi_n = E_n^2 \psi_n(x)
\label{sch}
\end{equation}
$\psi_n$ are required to be normalized.

Varying effective action (\ref{sgood}) with respect to $\la$  we get the first saddle point equation:
\begin{equation}
\cfrac{1}{2} \sum_n \cfrac{|\psi_n(x)|^2}{E_n} + |\si(x)|^2 - r =0 
\label{sp}
\end{equation}
To obtain this equation we have used the standard quantum mechanical first order perturbation theory for (\ref{sch}).

The second saddle-point equation coincides with the $\si$ equation of motion:
\begin{equation}
D_x^2 \si - \la(x) \si =0
\label{esi}
\end{equation}

Finally, we have to vary with respect to $A_x$:
\beq
\cfrac{i}{2} \sum_n \cfrac{\psi_n (D_x \psi_n)^* - \psi_n^* D_x \psi_n }{E_n} = i\si (D_x \si)^* - i\si_n^* D_x \si_n
\eeq
Below we will study the case $A_x=0$ with real $\psi_n$ and $\si$ and so this equation will be trivially satisfied.

\section{D-N boundary conditions: two phases}
Now it is time to choose boundary conditions. Let us consider the following:
For $N$ fields $n^i, i=1,\ldots,N$ we will use Dirichlet-Neumann (D-N):
\beq
n^i(0)=0, \ D_x n^i(L)=0
\eeq
And for $N$ fields $n^i,i=N+1,\ldots,2N$ we will use Neumann-Dirichlet(N-D):
\beq
D_x n^i(0)=0, \ n^i(L)=0
\eeq
And for $\si$ we will impose Neumann-Neumann(N-N):
\begin{equation}
D_x \si(0)=D_x \si(L)=0 
\label{}
\end{equation}
This choice breaks global $SU(2N+1)$ to $SU(N) \times SU(N)$.

Then in the D-N sector we have:
\beq
\psi_n(x)=\sqrt{\cfrac{2}{L}}\sin \l \cfrac{\pi x(n-1/2)}{L} \r,\ E_n^2=\l \cfrac{\pi (n-1/2)}{L}\r^2+\la,\ n=1,\ldots
\eeq
In the N-D sector:
\beq
\tilde \psi_n(x)=\sqrt{\cfrac{2}{L}} \cos \l \cfrac{\pi x(n-1/2)}{L} \r,\ E_n^2=\l \cfrac{\pi (n-1/2)}{L} \r^2+\la,\ n=1,\ldots
\eeq
If we plug this into the first saddle point equation (\ref{sp}) we will notice that $\sin^2$ and $\cos^2$ will sum up to $1$ and the $x$-dependence will disappear!
So we can consider $\si$ to be constant.  Let us first study the phase with non-zero $\la$. From the second saddle-point equation (\ref{esi}) we see
that we have to put $\si=0$. We will call this phase "Coulomb" phase because $n^i$ has zero VEV which leaves the $U(1)$ unbroken.

The first saddle-point equation now reads as:
\beq
\cfrac{N}{\pi} \sum_{n=1}^\infty \cfrac{1}{\sqrt{(n-1/2)^2+(\la L/\pi)^2}} - r =0 
\eeq

We need to separate the divergent part:
\beq
\cfrac{N}{\pi} \sum_{n=1}^\infty \l  \cfrac{1}{\sqrt{(n-1/2)^2+(\la L/\pi)^2}} - \cfrac{1}{n} \r + \cfrac{N}{\pi} \sum_{n=1}^\infty \cfrac{1}{n}   - r =0 
\eeq

Introducing the cut-off:
\beq
\sum_{n=1}^\infty \cfrac{\exp(-n \pi /L\La_\uv)}{n} = -\log(1-\exp(-\pi /L\La_\uv)) \approx -\log(\pi /L \La_\uv)
\eeq
Renormalizing $r$ using eq. (\ref{dyn}) we will have:
\beq
 \sum_{n=1}^\infty \l  \cfrac{1}{\sqrt{(n-1/2)^2+(\la L/\pi)^2}} - \cfrac{1}{n} \r   = \log(\pi/\La L)
\eeq
Now it is easy to see the presence of two phases: the maximum of the LHS is reached when $\la=0$, the corresponding value is
\beq
\sum_{n=1}^\infty \l \cfrac{1}{n-1/2}-\cfrac{1}{n} \r =\log(4) 
\eeq
It means that if $\log(\pi/\La L)>\log(4)$ the saddle-point equations do not have a solution
with non-zero $\la$.

Let consider the limit $L \ra 0$. We can expand the LHS in power series in $\la L$:
\beq
\cfrac{1}{\sqrt{(n-1/2)^2+(\la L/\pi)^2}} = \cfrac{1}{n-1/2} - 4 \l \cfrac{\la L}{\pi} \r^2 \cfrac{1}{(2n-1)^3}+\ldots 
\eeq
	
Using the following identity:
\beq
\sum_{n=1}^\infty \cfrac{4}{(2n-1)^3} = \cfrac{7}{2} \zeta(3)
\label{ids}
\eeq
we have:
\beq
\cfrac{7 \zeta(3)}{2} \l \cfrac{\la L}{\pi} \r^2 = \log(4 \La L/\pi)
\eeq
We see that the Coulomb phase does not exist for $L<\pi/4 \La$.

Let us now show that the "Higgs" phase $\si = \const, \la =0$ exists only for $L<\pi/4 \La$. We call this phase 
"Higgs" because non-zero $\si$ breaks $U(1)$ gauge symmetry. In this case the second saddle-point equation is
satisfied. The first one reads as:
\beq
\cfrac{N}{\pi} \sum_{n=1}^\infty \l  \cfrac{1}{n-1/2}  - \cfrac{1}{n} \r + \si^2   = \cfrac{N}{\pi}\log(\pi/\La L) 
\eeq
Again using eq. (\ref{ids}) we have:
\beq
\si^2=\cfrac{N}{\pi} \log(\pi/4 \La L)
\eeq

\section{D-D and N-N boundary conditions: one phase}

Instead of the D-N and N-D boundary conditions let us investigate the case with Dirichlet-Dirichlet(D-D) and Neumann-Neumann(N-N) boundary conditions.
As we will see shortly Coulomb phase is possible for all values of $L$.
For the D-D case we have the following set of eigenfunctions:
\beq
\psi_n(x)=\sqrt{\cfrac{2}{L}}\sin \l \cfrac{\pi x n}{L} \r,\ E_n^2=\l \cfrac{\pi n}{L}\r^2+\la,\ n=1,\ldots
\eeq
And for N-N:
\beq
\psi_n(x)=\sqrt{\cfrac{2}{L}}\cos \l \cfrac{\pi x n}{L} \r,\ E_n^2=\l \cfrac{\pi n}{L}\r^2+\la,\ n=0,\ldots
\eeq
Note that now we can have $n=0$ which corresponds to a constant mode. Note that if $\la=0$ we have a genuine zero mode. It means that
the phase with $\la=0$ can not exist for this choice of boundary conditions. 
In the saddle-point equations $\cos^2$ and $\sin^2$ again sum to 1, so we can have a saddle-point with constant $\si$ and $\la$. From now on,
we will assume that $\la=\const \neq 0$. Then from the second saddle-point equation it follows that $\si=0$.
The first saddle-point equation now reads as:
\beq
\cfrac{N}{\pi} \sum_{n=1}^\infty \l  \cfrac{1}{\sqrt{n^2+(\la L/\pi)^2}} - \cfrac{1}{n} \r + \cfrac{N}{\la L} + \cfrac{N}{\pi} 
\sum_{n=1}^\infty \cfrac{1}{n}   - r =0 
\eeq
After $r$ renormalization we have:
\beq
\cfrac{N}{\pi} \sum_{n=1}^\infty \l  \cfrac{1}{\sqrt{n^2+(\la L/\pi)^2}} - \cfrac{1}{n} \r + \cfrac{N}{\la L}  = \cfrac{N}{\pi} \log(\pi/\La L) 
\eeq
Unlike the D-N and N-D case now the LHS is not bounded from above because of the $\cfrac{N}{\la L}$ term, which is essentially the contribution from the
N-N constant mode. It easy to show that for a fixed $\La$ and $L$ we can always find the corresponding value of $\la$(for example one can plot the LHS as a function
of $\la$ and see that it takes
values from $-\infty$ to $+\infty$).

\section{Conclusion}
In this paper we studied the large $N$ $\cp^N$ model on a finite interval.  We have shown that for a specific choice of boundary conditions the saddle-point equations admit a simple analytical solution. Under the Dirichlet-Dirichlet and Neumann-Neumann boundary condition the system possesses a Coulomb phase with the uniform $\la$ VEV, usual for the $\cp^N$ in the infinite space.
This phase exists for all values of the interval length $L$. 
However, under the mixed Dirichlet-Neumann boundary conditions the system has two phases: Coulomb phase which
exists for $L>\pi/4\La$ and unusual Higgs phase for $L<\pi/4\La$ with the uniform $n^0$ VEV. Strictly speaking, it is possible
to have additional phases with non-constant VEVs, similar to the FFLO\cite{ff,lo} phase in superconductivity. It is even possible that the Coulomb and Higgs phases in the N-D case are not adjacent on the phase diagram because of the presence
of additional phases. We will postpone this analysis for future work.

\section*{Acknowledgment}
We are thankful to A. Gorsky for numerous discussions and F. Popov for reading the manuscript. A.M. is grateful to RFBR grant 15-02-02092 for travel support.
\printbibliography
\end{document}